# Quantitative Susceptibility Mapping using Deep Neural Network: QSMnet

Jaeyeon Yoon[1,†], Enhao Gong[2,†], Itthi Chatnuntawech[3], Berkin Bilgic[4], Jingu Lee[1], Woojin Jung[1], Jingyu Ko[1], Hosan Jung[1], Kawin Setsompop[4], Greg Zaharchuk[5], Eung Yeop Kim[6], John Pauly[2], and Jongho Lee[1,*]

[1]Laboratory for Imaging Science and Technology, Department of Electrical and Computer Engineering, Seoul National University, Seoul, Korea
[2]Department of Electrical Engineering, Stanford University, Stanford, California, USA
[3]National Nanotechnology Center, Pathum Thani, Thailand
[4]Department of Radiology, Harvard Medical School, Boston, MA, USA
[5]Department of Radiology, Stanford University, Stanford, California, USA
[6]Department of Radiology, Gachon University, Incheon, Korea

## ABSTRACT

Deep neural networks have demonstrated promising potential for the field of medical image reconstruction. In this work, an MRI reconstruction algorithm, which is referred to as quantitative susceptibility mapping (QSM), has been developed using a deep neural network in order to perform dipole deconvolution, which restores magnetic susceptibility source from an MRI field map. Previous approaches of QSM entail multiple orientation data (e.g. Calculation of Susceptibility through Multiple Orientation Sampling or COSMOS) or regularization terms (e.g. Truncated K-space Division or TKD; Morphology Enabled Dipole Inversion or MEDI) to solve the ill-conditioned deconvolution problem. Unfortunately, they either require long multiple orientation scans or suffer from artifacts. To overcome these shortcomings, a deep neural network, QSMnet, is constructed to generate a high quality susceptibility map from single orientation data. The network has a modified U-net structure and is trained using gold-standard COSMOS QSM maps. 25 datasets from 5 subjects (5 orientation each) were empoyed for patch-wise training after doubling the data using augmentation. Two additional datasets of 5 orientation data were used for validation and test (one dataset each). The QSMnet maps of the test dataset were compared with those from TKD and MEDI for image quality and consistency in multiple head orientations. Quantitative and qualitative image quality comparisons demonstrate that the QSMnet results have superior image quality to those of TKD or MEDI and have comparable image quality to those of COSMOS. Additionally, QSMnet maps reveal substantially better consistency across the multiple orientations than those from TKD or MEDI. As a preliminary application, the network was tested for two patients. The QSMnet maps showed similar lesion contrasts with those from MEDI, demonstrating potential for future applications.

## INTRODUCTION

Magnetic susceptibility is an intrinsic property of a tissue that determines the degree of magnetization in an external magnetic field. In recent years, measuring magnetic susceptibility using MRI has received much attention due to its potentials for the clinical diagnosis of diseases and the quantification of susceptibility sources (Duyn, 2013; Shmueli et al., 2009; Wang and Liu, 2015; Liu et al.,2015).

In MRI, a magnetic susceptibility source perturbs the main magnetic field and induces resonant frequency variation in and outside of the source. The spatial pattern of the resonant frequency variation has been demonstrated to be a dipole shape when a point susceptibility source is placed in the field (Salomir et al., 2003; Bowtell et al., 2003). For a more complex susceptibility source distribution, one can calculate a resonance frequency variation map or a field map by the convolution of the source distribution and the dipole pattern. The field map can be measured by the phase of a gradient-echo (GRE) sequence.

From the phase image of GRE, one can regenerate the susceptibility source distribution by performing the spatial de-convolution of the dipole pattern or, practically, by performing Fourier transforms of both phase image and dipole pattern and then dividing the Fourier transformed phase image by the Fourier transformed dipole pattern. This process of generating the susceptibility source distribution is referred to as quantitative susceptibility mapping (QSM). Unfortunately, the de-convolution or division in the Fourier domain is an ill-conditioned problem (Shmueli et al., 2009) because the Fourier transformed dipole pattern contains zeros (i.e. division by zero problem). A number of solutions have been proposed to address this reconstruction problem.

One approach is to truncate the inversion of the dipole pattern in k-space, avoiding infinity from the division by zero (Truncated K-space Division or TKD) (Shmueli et al., 2009). This approach generates a susceptibility map but it suffers from streaking artifacts. To overcome such artifacts, more sophisticated regularization algorithms that enforces certain prior information for QSM reconstruction have been proposed (Bilgic et al., 2014; de Rochefort et al., 2010; Li et al., 2015; Li et al., 2011; Liu et al., 2012; Liu et al., 2009; Schweser et al., 2012; Wharton et al., 2010). For example, the Morphology Enabled Dipole Inversion (MEDI) algorithm (Liu et al., 2011) constrains the reconstructed map to have similar edges to the $T_2^*$-weighted magnitude image. The results show much reduced streaking artifacts and a refined QSM map. However, certain errors from the ill-conditioned problem still remain (Liu et al., 2012). When prolonged scan time is not a limitation and a subject is cooperative, one can obtain a more accurate QSM map by acquiring GRE data from multiple head orientations with respect to $B_0$ and reconstructing the data using the Calculation of Susceptibility through Multiple Orientation Sampling (COSMOS) algorithm (Liu et al., 2009; Wharton and Bowtell, 2010). The resulting QSM map can be considered as a gold-standard when ignoring susceptibility and structural anisotropies in white matter (Lee et al., 2017; Lee et al., 2010; Liu, 2010; Wharton and Bowtell, 2012). However, this approach requires the acquisition of multiple head orientations and is not practical for a clinical routine.

Recently, deep learning using a neural network has been widely applied across multiple fields including computer vision, computer assisted diagnosis, pattern recognition (Xie et al.,2012; Noh et al., 2015). A deep neural network has shown the capability of performing a non-linear mapping from an input space to an output space. When training dataset is large enough, the deep neural network outperformed the state-of-art machine learning algorithms or even human raters (Mao X et al., 2016; Kim J., 2016; He etl al., 2015). Deep neural networks have also been applied for medical image reconstruction such as X-ray, CT, PET, and MRI (Prasoon et al., 2013; Teramoto et al., 2016; Yang et al., 2018; Zichuan et al., 2017; Park et al., 2016). The methods have showed impressive performances compared to conventional algorithms. Considering current outcomes of the deep neural network for image reconstruction, the approach may be applicable for the QSM reconstruction. In this paper, we explore the possibility of training a deep neural network to conduct the dipole deconvolution task for high quality reconstruction of QSM. We will refer this neural network as QSMnet hereafter. The target reconstruction quality of QSMnet is that of COSMOS QSM but only a single head orientation image is used for the QSM reconstruction of the neural network.

*Corresponding author: Jongho Lee, Ph.D, Building 301, Room 1008, 1 Gwanak-ro, Gwanak-gu, Seoul, Korea, E-mail: jonghoyi@snu.ac.kr

†Co-first authors: These authors equally contributed to the paper

## MATERIALS AND METHODS

### MRI data acquisition and processing

For the training and testing of QSMnet, total 35 scans from seven healthy volunteers were acquired at five different head orientations per volunteer (at 3T; six datasets using Tim Trio, and three datasets using Skyra, SIEMENS, Erlangen, Germany). For the evaluation of the neural network, two patients were scanned at a single head orientation (3T; Skyra, SIEMENS, Erlangen, Germany). All subjects signed IRB-approved written consent forms.

After three plane localization and shimming, a 3D single-echo GRE scan was acquired using following sequence parameters: FOV = 256 x 224 x 176 $mm^3$ (224 x 224 x 176 $mm^3$ for healthy volunteers at Skyra; 192 x 192 x 80 $mm^3$ for patients), voxel size = 1 x 1 x 1 $mm^3$, TR = 33 ms, TE = 25 ms, bandwidth = 100 Hz/pixel, flip angle =15°, acceleration factor = 2 x 2 (acceleration factor = 2 for patients), and total acquisition time = 5 min 46 sec (5 min 18 sec for patients). For healthy volunteers, the scan was repeated five times with an instruction of changing head orientation after each scan. Before each scan, manual shimming was performed to improve field homogeneity. All GRE scans were acquired axially. When subject motion was observed, the scan was repeated.

The magnitude and phase images were reconstructed from k-space data using offline GRAPPA reconstruction (Griswold et al., 2002) followed by coil combination using sensitivities estimated with ESPIRiT (Uecker et al., 2014). From the magnitude image, a brain mask was generated using BET (FSL, FMRIB, Oxford, UK) (Smith, 2002). Within the mask, the phase image was spatially unwrapped by Laplacian phase unwrapping (Li et al., 2011). Then a local field map was generated by removing background field using V-SHARP (Wu et al., 2012).

For the COSMOS reconstruction of the multiple head orientation data, the local field maps from the five different head orientations were registered as follows: First, each orientation magnitude image was registered to the unrotated head orientation magnitude image to calculate a rotation matrix (FSL's FLIRT) (Jenkinson et al., 2002; Jenkinson and Smith, 2001). Then, the rotation matrix was applied to the local field map for the registration. Using the registered local field map and the rotation information, a QSM map was generated using the COSMOS algorithm (Liu et al., 2009).

In addition to the COSMOS QSM results, single orientation QSM maps were generated using the TKD and MEDI algorithms. For TKD, the filter truncation value was set to 5 (Shmueli et al., 2009). For MEDI, the regularization factor was set to 3000 (Liu et al., 2011). All five head orientation phase images were reconstructed, generating five QSM maps that were expected to show consistent contrasts across the head orientations. The patient data were also processed with TKD and MEDI.

### Deep neural network for QSM: QSMnet

Out of the seven healthy volunteer datasets of the multiple head orientations, 25 scans from five subject datasets were used for the training of QSMnet, 5 scans from one subject dataset was used as a validation set, and 5 scans from one subject dataset was used as a test set.

The expected function of QSMnet is to perform the deconvolution of the dipole pattern. In order to train consistent deconvolution for our multiple head orientation datasets, the unregistered phase images, that had $B_0$ along z-axis, were applied as the input of QSMnet. Then, a "rotated" COSMOS QSM map that matched to the orientation of the input phase image was used as the label of QSMnet (see Figure S1 in supplementary information). The rotated COSMOS QSM map was generated by matrix rotation using a transpose of the rotation matrix, which was calculated during the magnitude registration.

To increase the size of the training datasets, the COSMOS QSM maps were rotated in an angle (from $-30^0$ to $+30^0$ relative to $B_0$) and local field maps were generated by dipole convolution. Using this data augmentation process, the total training datasets were doubled and total 50 scans were used for training.

Since the dipole deconvolution, that relates local field to susceptibility, is defined in 3D, a 3D patch with the size of 64 x 64 x 64 voxels was used for training. The patch was generated with an overlapping scheme of 66% overlap between adjacent patches. The total number of patches for training was 16,800.

In our QSMnet reconstruction, the input (local field image) and the output (QSM map) share similar structures and have the same matrix size. Hence, a U-net, which was proposed for biomedical image segmentation (Ronneberger et al., 2015), was used as a base structure of QSMnet. The network structure was modified from 2D to 3D to train a 3D dipole deconvolution (Figure 1). The network consisted of 19 convolutional layers, 18 batch normalization, 18 ReLU nonlinear layers, 4 max-pooling layers, 4 deconvolution layers, and 4 feature contracting paths. The first half of the network had four groups and each group contained two sets of convolutional layers with a 5 x 5 x 5 kernel, batch normalization, and ReLU. Each group was connected by a max-pooling layer. The second half of the network had four groups, containing additional feature concatenation layers compared to the first half groups. Each group was connected by a deconvolution layer instead of the max-pooling layer. Two groups were connected by two convolutional layers. Finally, the last layer applied a 1 x 1 x 1 convolution kernel. The numbers of channels of convolutional layers are summarized at the bottom of blocks in Figure 1.

Three loss functions were designed to incorporate physical model consistency (Model loss), voxel-wise difference (L1 loss), and image edge preservation (Gradient loss).

For Model loss, the L1 difference between the dipole convolution of the label and the output was measured as follows:

$$\text{loss}_{Model} = ||d * \chi - d * y||_1 \quad [\text{Eq. 1}]$$

where $d$ is the dipole kernel, $\chi$ is the output, and $y$ is the label (i.e. COSMOS QSM). This loss enforces the consistency in the dipole model between the label and the output. To avoid incorrect values at the edges from the convolution process, 5 voxels at the edges were discarded.

The L1 loss was defined as the L1 difference between the label and the output of QSMnet, quantifying the voxel difference.

$$\text{loss}_{L1} = ||\chi - y||_1. \quad [\text{Eq. 2}]$$

The third loss was the gradient difference loss to preserve edge information in the reconstructed map (Mathieu et al., 2016).

$$\text{loss}_{Gradient} = ||\nabla\chi|_x - |\nabla y|_x| + ||\nabla\chi|_y - |\nabla y|_y| + ||\nabla\chi|_z - |\nabla y|_z| +$$
$$||\nabla(d*\chi)|_x - |\nabla(d*y)|_x| + ||\nabla(d*\chi)|_y - |\nabla(d*y)|_y| +$$
$$||\nabla(d*\chi)|_z - |\nabla(d*y)|_z|. \quad [\text{Eq. 3}]$$

The total loss was the weighted sum of the three losses.

$$Total\ loss = w_1 \text{loss}_{Model} + w_2 \text{loss}_{L1} + w_3 \text{loss}_{Gradient} \quad [\text{Eq. 4}]$$

where the weights were empirically determined as $w_1 = 1$, $w_2 = 1$, and $w_3 = 0.1$. The minimization was performed with RMSProp Optimizer. The learning rate was exponentially decayed from 0.001 at every 400 steps of the training. The initial values of convolutional kernel were calculated by Xavier initializer (Glorot and Bengio, 2010; Jia et al., 2014). The batch size was set to be 12 and training was stopped at 50 epochs as the performance was shown to be stable. The network was trained and evaluated using TensorFlow (Rampasek and Goldenberg, 2016) using NVIDIA 1080TI GPU.

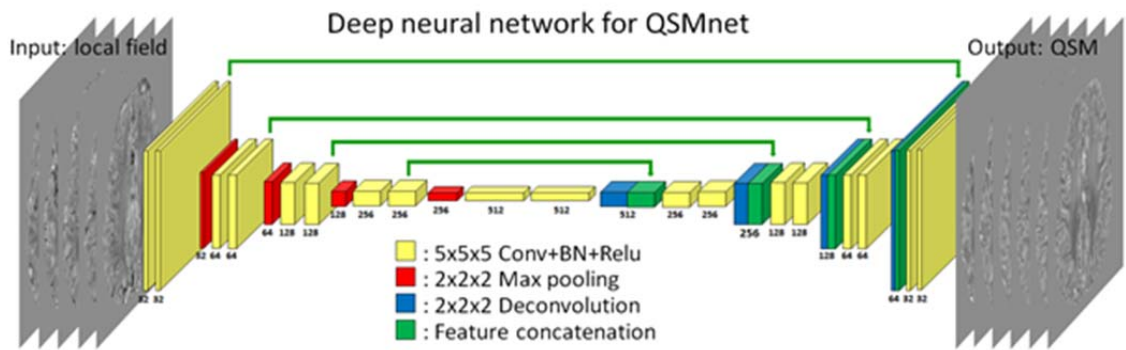

Figure 1. Network structure of QSMnet. A 3D U-net was designed with 18 convolutional layers (kernel size = 5 x 5 x 5), 1 convolutional layer (kernel size = 1 x 1 x 1), 4 max pooling layer strides (kernel size = 2 x 2 x 2), 4 deconvolution layer strides (kernel size = 2 x 2 x 2), and 4 feature concatenations.

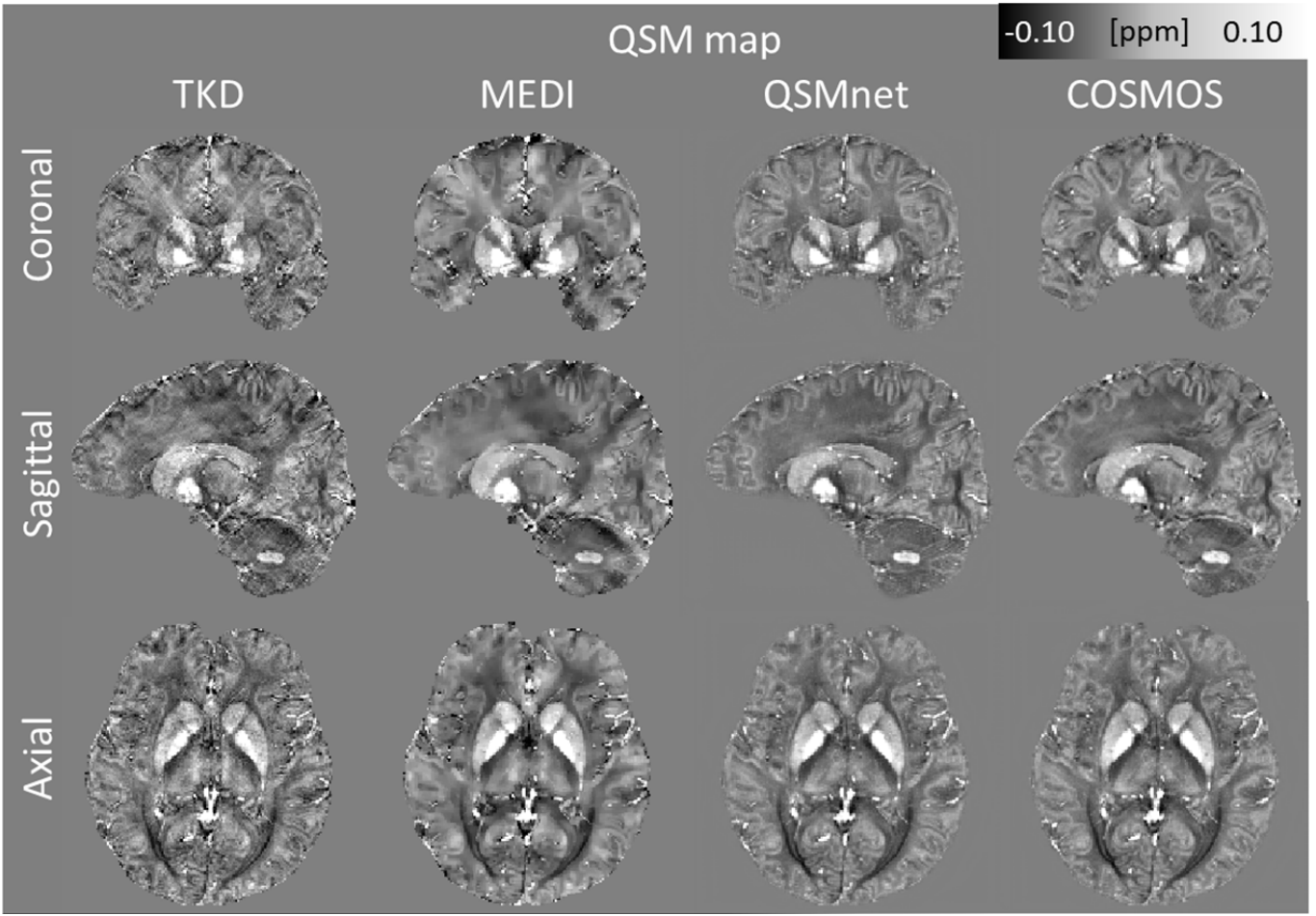

Figure 2. QSM maps of the test set reconstructed by the four methods. The QSMnet map has high fidelity to the gold-standard COSMOS map. On the other hand, the TKD and MEDI maps shows steaking artifacts. See Figure S3 in supplementary information for residual error maps.

**Evaluation of QSM algorithms**

To compare image quality of the QSM maps, the test set with the five head orientations were processed using the three reconstruction methods. The reconstruction times for QSMnet and MEDI were measured. The reconstructed QSM maps from the methods were displayed in three plains. The residual error maps with respect to the gold-standard COSMOS QSM map were generated. The quantitative metrics, peak signal-to-noise ratio (pSNR), root mean squared error (RMSE), high-frequency error norm (HFEN), and structure similarity index (SSIM), that were used to measure the reconstruction quality of a QSM algorithm (Langkammer et al., 2018) were calculated with the COSMOS QSM map as a reference. The means and standard deviations of the metrics for the five head orientations were compared.

To quantify the accuracy and consistency of the QSM maps from the multiple head orientations, a region of interest (ROI) analysis was performed. Five ROIs, putamen (PUT), globus pallidus (GP), caudate (CAU), red nucleus (RN), and substantia nigra (SN), were manually segmented (see supplementary information; Figure S2). For each ROI, the mean and standard deviation was calculated from the five head orientation QSM maps.

As a preliminary attempt to explore the applicability of QSMnet, the two patients, one with a microbleed and the other with multiple sclerosis lesions, which have not been trained in the network, were reconstructed and compared the results with those from MEDI.

**RESULTS**

The three plain views of the QSM maps from the test set are displayed in Figure 2. The residual error maps are included in supplementary information (Figure S3). These maps are from the first scan that had neutral head orientation (i.e. scan with no instruction of head tilting). The coronal and sagittal views of the TKD and MEDI results (first and second columns, respectively) depict streaking artifacts that deteriorate the image quality. On the other hand, the QSM map from QSMnet (third column) shows no noticeable artifacts. When compared to the COSMOS result (last column), the QSMnet map reveals almost identical contrast. The residual error maps also confirm these observations (Figure S3). The residual errors are the lowest in the QSMnet map (RMSE errors: 0.034 for TKD, 0.029 for MEDI, 0.016 for QSMnet).

The quantitative metrics, pSNR, RMSE, HFEN, and SSIM, of the three reconstruction methods are summarized in Table 1. Compared to TKD or MEDI, the QSMnet results achieved the highest pSNR (the higher the better), lowest RMSE (the lower the better), lowest HFEN (the lower the better), and highest SSIM (the higher the better), suggesting the best performances for all criteria.

The QSM maps from the five head orientations are presented in Figure 3. Each row shows the QSM maps reconstructed by TKD, MEDI, and QSMnet, respectively. In the last column, the COSMOS QSM map is shown as a reference. The QSMnet results reveal excellent fidelity to the COSMOS map for all head orientations. When compared to the maps from the other methods, the QSMnet results reveal superior consistency across the head orientations. In particular, both TKD and MEDI maps suffer from streaking artifacts as indicated by the red arrows, but the QSMnet maps do not show the artifacts. The green arrows indicate internal capsule that changes the contrast in different head orientations in the TKD and MEDI results. On the other hand, the QSMnet maps reveal consistent results. The axial and sagittal views of the QSMnet maps (see Figure S4 in supplementary information) reconfirm superior image quality with unnoticeable steaking artifacts and higher consistency across the head orientations.

Table 1. Quantitative performance metric, pSNR, RMSE, HFEN, and SSIM, from the three reconstruction methods. QSMnet shows the best performances in all criteria.

|  | pSNR(dB) | RMSE(%) | HFEN(%) | SSIM |
| --- | --- | --- | --- | --- |
| TKD | 38.1±0.5 | 88.7±4.9 | 81.4±8.6 | 0.80±0.05 |
| MEDI | 39.1±0.3 | 85.6±4.9 | 83.6±7.6 | 0.86±0.04 |
| QSMnet | 42.0±0.9 | 57.1±6.0 | 55.6±7.0 | 0.90±0.03 |

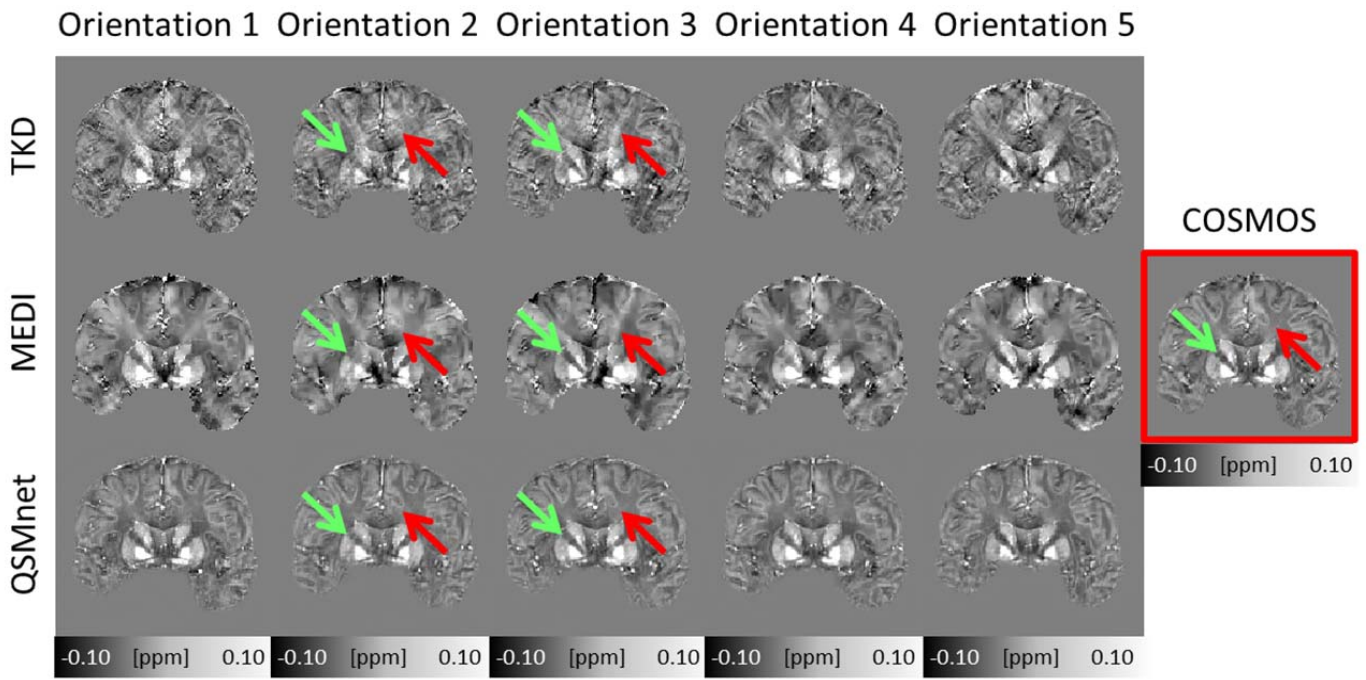

Figure 3. Coronal view of the QSM maps from the five head orientations. Compared to the TKD and MEDI results, the QSMnet maps show consistent results across the orientations. The QSMnet maps are close to that of COSMOS. Red arrows indicate streaking artifacts in the TKD and MEDI maps. The artifacts are not present in the QSMnet map. Green arrows point internal capsule that reveals a consistent contrast across the orientations in QSMnet. On the other hand, the structure shows substantial contrast variations in the MEDI and TKD results. See Figure S4 in supplementary information for the axial and sagittal views.

Zoomed-in views of Figure 3 are displayed in Figure 4. The blue circles contain cortical ribbons of the cortex. As compared to the other methods, the QSMnet results preserves detail structures for the five head orientations, suggesting potential for applying the method for cortical imaging.

When the mean susceptibility and standard deviation across the head orientations are calculated in the ROIs (Figure 5; TKD: pink, MEDI: blue, QSMnet: red, COSMOS: green), the QSMnet results show the tightest error bars with the smallest error when compared to the gold-standard results of COSMOS. This result demonstrates the superior accuracy of the QSMnet when compared to the other reconstruction methods.

Another advantage of the QSMnet is reconstruction speed. The average reconstruction time was only 6.3 ± 0.0 sec. and was much faster than MEDI (255.8 ± 18.2 sec.).

When QSMnet was applied to the patients with a microbleed or multiple sclerosis lesions (Figure 6), the results reveal comparable contrasts to those of MEDI.

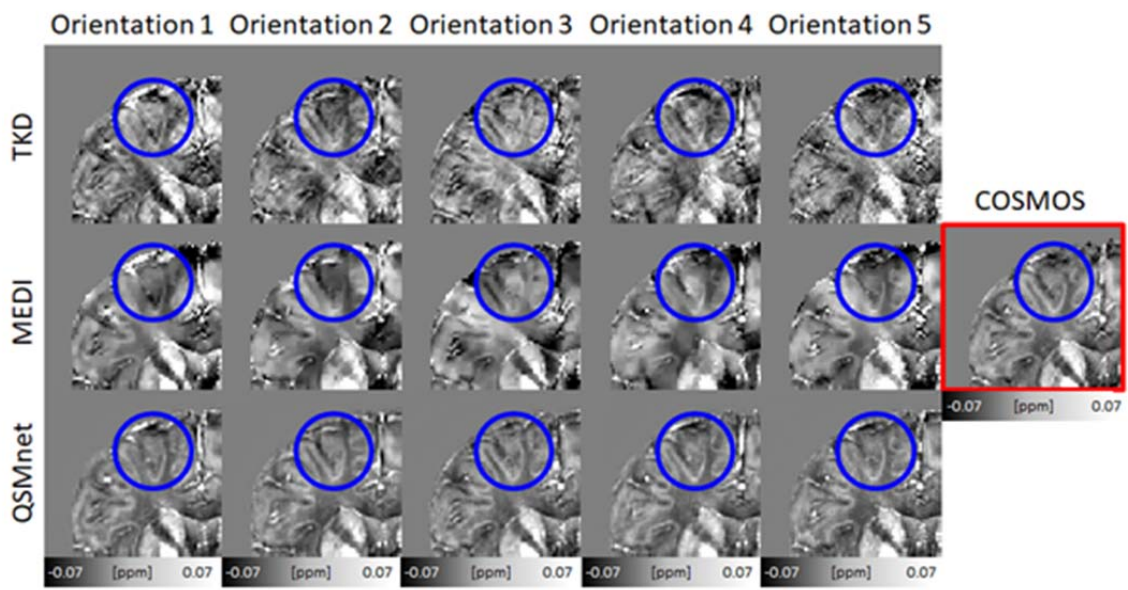

Figure 4. Zoomed-in view of Figure 3. The blue circles encompass cortical ribbons in cortex. The QSMnet maps preserve detail structures of the cortical ribbon while TKD and MEDI results lose information.

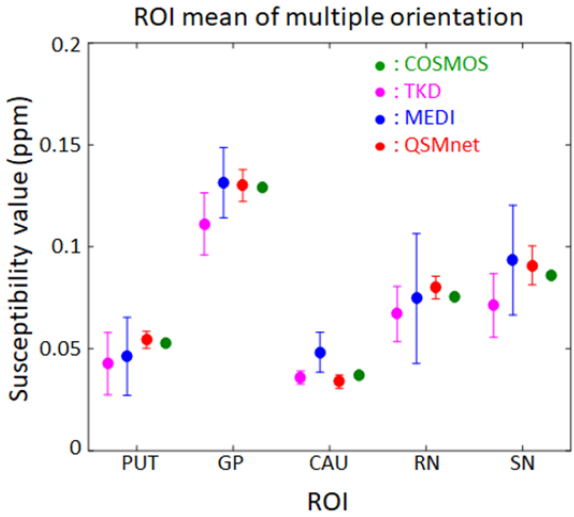

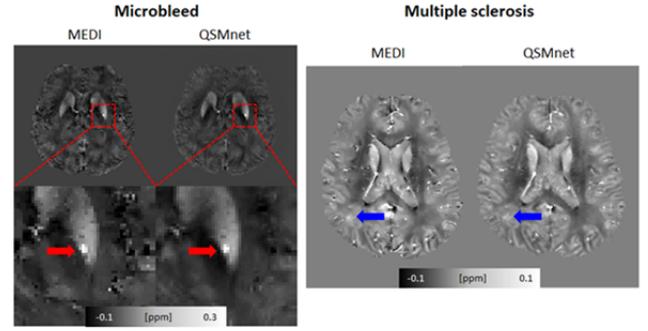

Figure 5. ROI analysis of the five head orientations. When the susceptibility values of the ROIs (PUT, GP, CAU, RN, and SN) are plotted, the QSMnet results match well with the gold-standard COSMOS QSM results with the smallest standard deviation (green for COSMOS, pink for TKD, blue for MEDI, and red for QSMnet).

Figure 6. QSM maps from the patients with microbleed (left, red arrows) and multiple sclerosis lesions (right, blue arrows) are compared using MEDI and QSMnet. The lesions are similarly delineated in both maps. The multiple sclerosis lesions show more positive susceptibility contrast presumably due to demyelination. They are better identified when compared with negative contrast in the contralateral side white matter. Note that no lesion was observed in the healthy volunteers and, therefore, was trained in QSMnet.

## DISCUSSION AND CONCLUSION

In this work, we constructed a deep neural network (QSMnet) to perform the dipole deconvolution in the QSM reconstruction. The results demonstrated high quality QSM maps that were close to the gold-standard COSMOS QSM maps. In particular, the QSM maps from our network demonstrated consistent magnetic susceptibility results for multiple head orientation inputs. This outcome has an important value since it supports high reproducibility for a longitudinal study that requires multiple scans over time. The preliminary patient results suggest that the network may be applied for abnormality that has not been trained for the network although the patient study was limited to only two patients. Further validation using a large number of patients with different diseases and lesion types is warranted to confirm the applicability of the neural network for patient studies.

Despite its high expressive power, one of the biggest challenges of a neural network is that it is difficult to characterize. Since the network is trained by data with no specific design for a target function, it is difficult to understand how it functions and is difficult to interpret the outcomes. Hence, reliability, which is one of the most valuable criteria for clinical applications, is difficult to guarantee. Therefore, our demonstration of the QSM results needs to be interpreted with caution. In our QSMnet, the loss function was designed with a term ($loss_{Model}$) that enforces the dipole deconvolution function. This approach may have helped the network to learn the physical model rather than image to image transformation. However, there is still no guarantee on the training result. Despite this concern, the outcomes of the patients are encouraging since these features (e.g. microbleed and multiple sclerosis lesions) are not included in the training sets but they are correctly reconstructed by the network. We, therefore, speculate that the network is trained for the dipole deconvolution function. Further exploration is necessary to understand the characteristics of the network.

When the multiple head orientation QSM maps were carefully examined, we observed that white matter showed consistent contrasts in the QSMnet results. This result suggests the network provides high reproducibility but it may not be correct physically due to the anisotropic characteristics of white matter susceptibility and microstructure (Lee et al., 2010; Oh et al., 2013; Wharton and Bowtell, 2012). One potential interpretation of the observation is that the neural network suppresses the anisotropy. This suppression of anisotropy is expected since the network was trained only for isotropic susceptibility (i.e. COSMOS). An alternative interpretation is that the effects of the anisotropy were small since the average head orientation was only 16.9º $\pm$ 4.0º. We believe both explanations have contributed to the results. If data with more head orientations were available for training, one may train a neural network for susceptibility tensor imaging (Liu, 2010), generating more accurate susceptibility results.

For MEDI, the regularization factor affects the image quality. When multiple regularization factors were tested the results still show large variability in QSM results for the multiple head orientations (see Figure S5 in Supplementary Information).

The proposed neural network has limitations. First, the input resolution is fixed. For lower resolution data than the trained resolution, one may interpolate the data before processing. For higher resolution data, however, the reconstruction may not be properly performed (Shmueli et al., 2016). A new network with new datasets may be necessary for higher resolution data. For different matrix size data (i.e. same resolution but different FOV), our network can still process the results since the convolutional layer conducts their operation sequentially regardless of the input matrix size. Our network presumes the z-axis of the input image as the $B_0$ field direction. If input data have a different orientation, a matrix rotation needs to be applied to match the $B_0$ direction.

The proposed QSMnet will need to be tested extensively for both healthy volunteers and patients to confirm the validity of the method. Additionally, efforts to understand the characteristics of the network will be continued. The advantages of the high quality map, which is close to gold-standard, and high processing speed, which is close to real-time, may provide the method a valuable opportunity for future applications of the method.

## ACKNOWELDGEMENTS

This research was supported by National Research Foundation of Korea (NRF-2017M3C7A1047864) and Creative-Pioneering Researchers Program of Seoul National University.

# SUPPLEMENTARY INFORMAITON

**Supplementary information 1. Data processing pipeline**

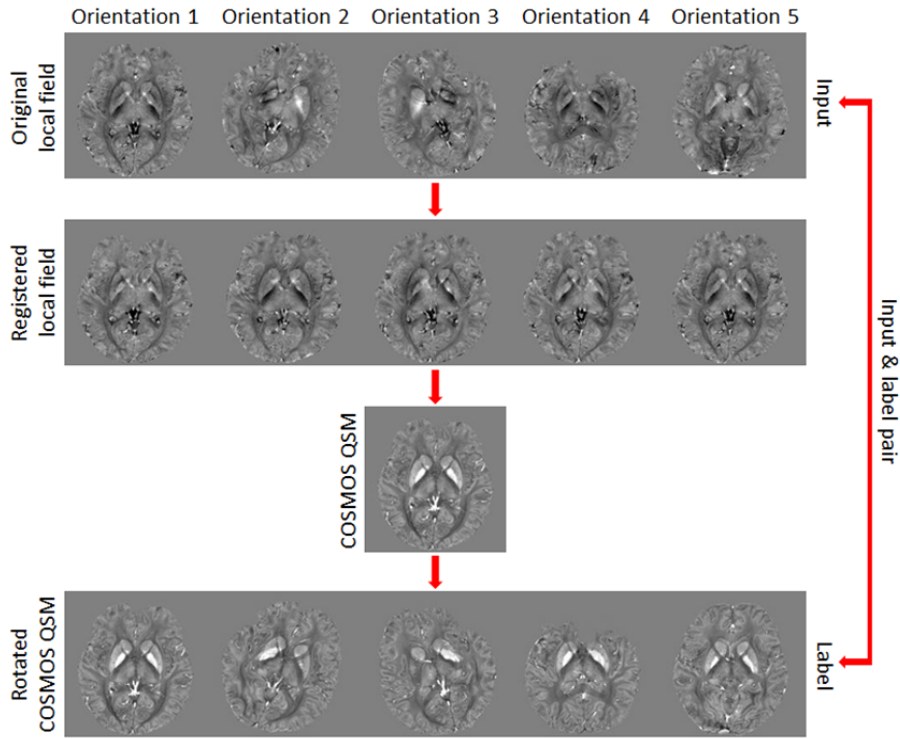

Figure S1. First row shows original local field maps from the scans. These maps were registered to the first scan generating the registered local field map (second row). From the registered local field map, the COSMOS QSM map is computed (third row). This map was rotated back to the original head orientation (forth row) to generate input and label pairs for QSMnet training.

**Supplementary information 2. Regions of interest**

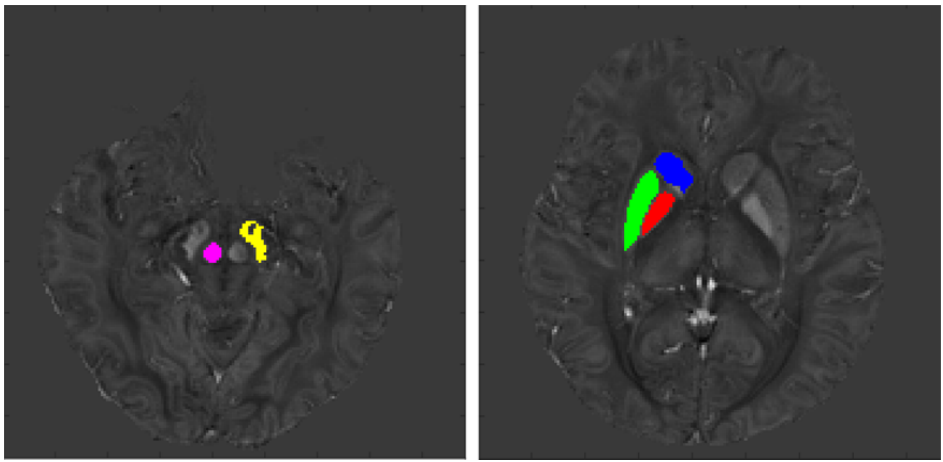

Figure S2. Five ROIs: Red nucleus (pink) and substantia nigra (yellow) in the left figure, and globus pallidus (red), putamen (green), and caudate (blue) in the right figure.

**Supplementary information 3. Residual error maps**

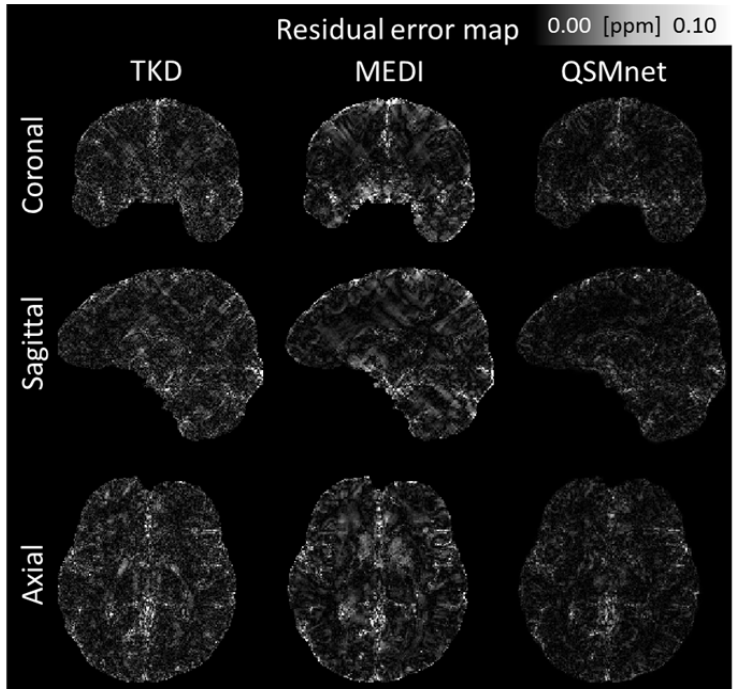

Figure S3. The residual error maps of Figure 2 referenced to the COSMOS result. Large errors are observed in the TKD and MEDI results. Streaking artifacts are more pronounced in the TKD and MEDI maps.

**Supplementary information 4. Axial and sagittal views of the QSM maps from the five head orientations**

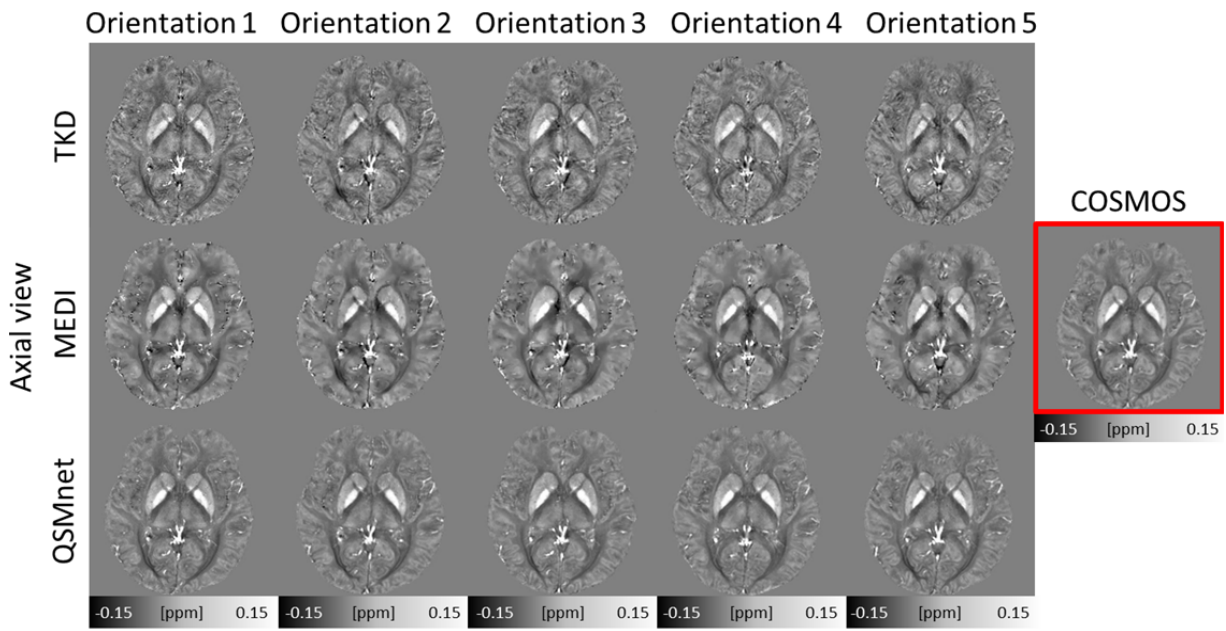

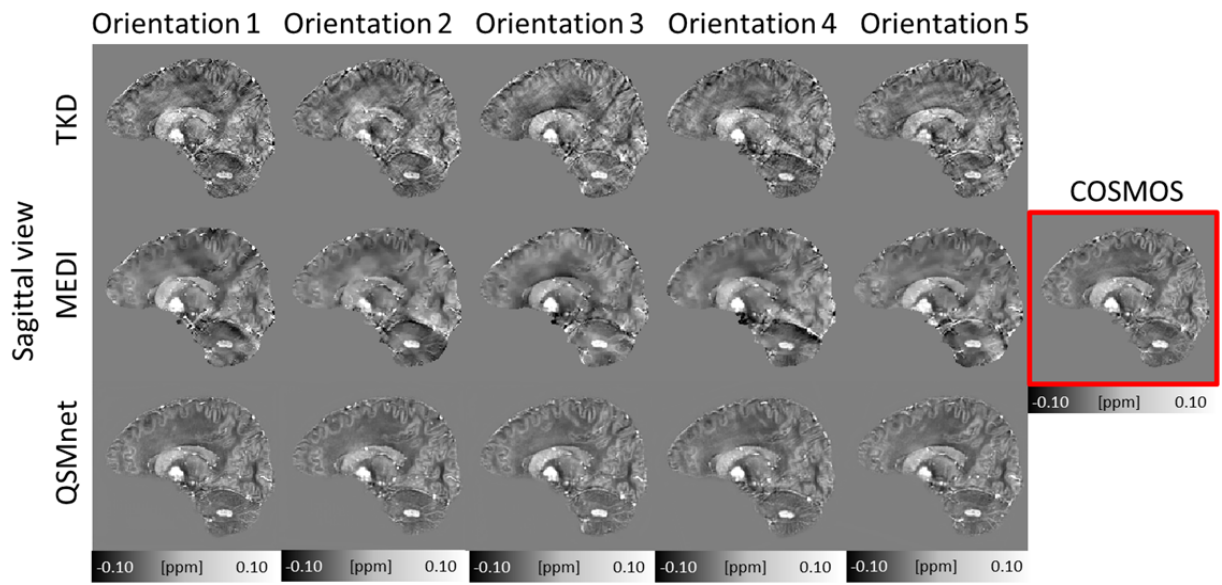

Figure S4. Axial and sagittal views of the QSM maps from the three reconstruction methods (TKD, MEDI and QSMnet in each row). The COSMOS QSM maps are shown in the last column. Both TKD and MEDI results contain streaking artifacts, creating variability in the different orientation maps. On the other hand, in QSMnet, no streaking artifacts is visually detected in all head orientations.

**Supplementary information 5. Effects of the regularization factor in MEDI**

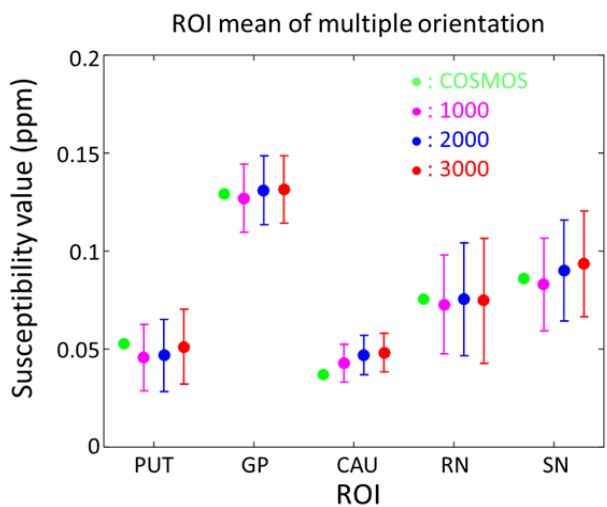

Figure S5. For MEDI, three regularization factors ($\lambda = 1000$, 2000 and 3000) were tested. For the smallest regularization, substantial smoothing was observed and cortical structures were difficult to observed. When the value was increased to 3000, better delineation of structures was detected at the cost of increased streaking artifacts. When the three values applied for the ROI analysis, the results show persistent variability for the multiple head orientations. Figure shows the means and standard deviations of the five orientations for each regularization value and ROIs (green for COSMOS, pink for $\lambda = 1000$, blue for $\lambda = 2000$, and red for $\lambda = 3000$).